\documentclass[letterpaper, 10 pt, conference]{ieeeconf}  % Comment this line out if you need a4paper

\IEEEoverridecommandlockouts                              % This command is only needed if 
\usepackage{cite}
\usepackage{amsmath,amssymb,amsfonts}
\usepackage{diagbox}
\usepackage[noend]{algpseudocode}
\usepackage{algorithmicx,algorithm}
\usepackage{graphicx}
\usepackage{textcomp}
\usepackage{xcolor}
\usepackage{float}
\usepackage{subfigure}

\title{\LARGE \bf
OIDM: An Observability-based Intelligent Distributed Edge Sensing Method for Industrial Cyber-Physical Systems
}

\author{Shigeng Wang, Tiankai Jin, Yehan Ma and Cailian Chen
\thanks{The authors are with the Department of Automation, Shanghai Jiao Tong University, and Key Laboratory of System Control and Information Processing, Ministry of Education of China, Shanghai 200240, China. Emails: {wangshigeng, tiankaijin, yehanma, cailianchen}@sjtu.edu.cn.}%
}

\begin{document}

\maketitle
\thispagestyle{empty}
\pagestyle{empty}

%%%%%%%%%%%%%%%%%%%%%%%%%%%%%%%%%%%%%%%%%%%%%%%%%%%%%%%%%%%%%%%%%%%%%%%%%%%%%%%%
\begin{abstract}

Industrial cyber-physical systems (ICPS) integrate physical processes with computational and communication technologies in industrial settings. With the support of edge computing technology, it is feasible to schedule large-scale sensors for efficient distributed sensing. In the sensing process, observability is the key to obtaining complete system states, and stochastic scheduling is more suitable considering uncertain factors in wireless communication. However, existing works have limited research on observability in stochastic scheduling. Targeting this issue, we propose an observability-based intelligent distributed edge sensing method (OIDM). Deep reinforcement learning (DRL) methods are adopted to optimize sensing accuracy and power efficiency. Based on the system's ability to achieve observability, we establish a bridge between observability and the number of successful sensor transmissions. Novel linear approximations of observability criteria are provided, and probabilistic bounds on observability are derived. Furthermore, these bounds guide the design of action space to achieve a probabilistic observability guarantee in stochastic scheduling. Finally, our proposed method is applied to the estimation of slab temperature in industrial hot rolling process, and simulation results validate its effectiveness.

\end{abstract}

%%%%%%%%%%%%%%%%%%%%%%%%%%%%%%%%%%%%%%%%%%%%%%%%%%%%%%%%%%%%%%%%%%%%%%%%%%%%%%%%
\section{Introduction}

In the era of Industry 4.0 \cite{z1}, characterized by digitization, networking, and intelligence, an increasing integration of information technology and operational technology forms the industrial cyber-physical systems (ICPS). However, the influx of massive data and frequent transmissions poses increasingly complex scheduling challenges for ICPS. In this trend, a new generation of ICPS supported by the edge computing technology is developing rapidly \cite{z2}. Edge computing units (ECUs) are embedded closer to industrial production lines in order to enhance local computational capabilities, which reduces network congestion and energy consumption. Multiple ECUs collaborate to accomplish distributed edge sensing tasks, thereby facilitating efficient sensing in ICPS.

The sensing process not only directly affects the accuracy of system state estimation but also has a further impact on the design of control strategies. Thus, a considerable amount of existing research focuses on how to schedule the sensing devices to improve sensing performances, including sensing accuracy and power efficiency. Among these works, stochastic scheduling \cite{z3} is more suitable for real-world industrial scenarios than deterministic scheduling \cite{z4}, considering packet loss in wireless communication and the reliability of various system components. 
However, stochastic scheduling cannot strictly guarantee observability which is the key to obtaining complete system states. Additionally, it is difficult to handle observability criteria expressed in expectation. Some works \cite{z5}, \cite{z6} focus on deterministic methods guaranteeing observability, while observability analysis in stochastic scheduling remains an underexplored area. 

% Additionally, due to adopting probabilistic access of sensing devices, stochastic scheduling optimizes objectives in terms of expected values, potentially finding solutions that are superior to deterministic scheduling among a wider range of feasible solutions. 

Deep reinforcement learning (DRL) methods are emerging as effective tools for stochastic sensor scheduling. Deep Q-network (DQN) is applied to the wireless sensor scheduling in cyber-physical systems \cite{z7} and transmission scheduling for edge sensing \cite{z8}. In \cite{z9}, deep deterministic policy gradient (DDPG) is applied to stochastic sensor scheduling for remote estimation with channel capacity constraint. These studies model the stochastic scheduling as a Markov decision process (MDP) and search for the optimal solution based on a feasible one. 
Achieving observability is a sufficient condition for guaranteeing stability of the estimation error covariance. A policy with a probabilistic observability guarantee can also guarantee corresponding sensing performance, providing a feasible solution for optimization. 

% Stability of the estimation error covariance (EEC) plays a crucial role in the sensing process, and a policy guaranteeing stability used to be chosen as the feasible solution to start optimization. 

In industrial scenarios, some system states cannot be directly observed due to certain constraints. For example, in the hot rolling process, sensing devices can only observe the temperature on the surface of the slab \cite{z14}. Besides, the expanding network scale poses challenges to achieving observability. Thus, it is of practical importance to design sensor scheduling methods considering the probabilistic observability guarantee. 

Addressing the aforementioned issues, we combine DRL with observability analysis. The main contributions of this paper are summarized as follows:

\begin{itemize}
% 多ECU
\item New linear approximations of observability criteria for Multi-ECU ICPS are provided, and bounds on the observability probability are derived. 
\item A novel observability-based intelligent distributed edge sensing method (OIDM) is proposed to balance the sensing accuracy and power efficiency guaranteeing probabilistic observability.
\item OIDM is applied to estimate the slab temperature in a typical ICPS of industrial hot rolling, and its effectiveness is fully verified by simulation results.
\end{itemize}

The remainder of this paper is organized as follows. In Section \uppercase\expandafter{\romannumeral2}, the preliminaries and problem formulation are provided. In Section \uppercase\expandafter{\romannumeral3}, the probabilistic observability is explored, and OIDM is described in detail. In Section \uppercase\expandafter{\romannumeral4}, OIDM is applied to estimate the slab temperature in the hot rolling process. Finally, Section \uppercase\expandafter{\romannumeral5} concludes this paper.

\textbf{Notations:} The $(i,j)$-th entry of a matrix $C$ is $[C]_{i,j}$, and the $i$-th entry of a vector $x$ is $[x]_i$. The $i$-th row of a matrix $D$ is $D_{(i,:)}$. The main diagonal operator is denoted as diag[$\cdot$], the transpose as superscript T, the mathematical expectation as $\mathbb{E}[\cdot]$, the trace operator as tr($\cdot$), the probability as $\mathbb{P}[\cdot]$, the Kronecker product as $\otimes$, and the complement to set $\Omega$ as $\bar\Omega$.

\section{Preliminaries and Problem Formulation}

\subsection{Edge Computing Supported ICPS}

\begin{figure}[htbp]
\centerline{\includegraphics[width=8.2cm]{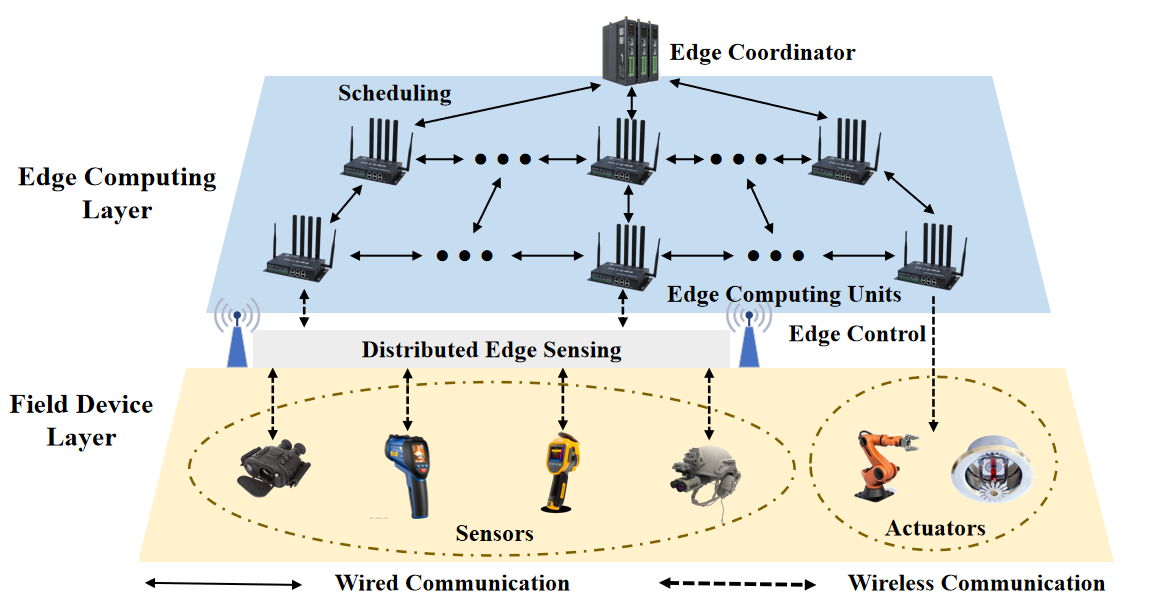}}
\caption{The architecture of edge computing supported ICPS.}
\label{fig:1}
\end{figure}

As shown in Fig. \ref{fig:1}, the edge computing supported ICPS establish a dual-layer architecture that facilitates collaboration between the edge and the endpoint. 

In the field device layer, numerous sensing devices are deployed to observe states of the production line. These sensing devices carry out sensing scheduling instructions issued by ECUs and transmit observations to the next higher layer wirelessly. 

After information collection from linked sensors, each ECU shares results of preprocessing with neighboring ECUs in the edge computing layer. Based on information fusion, each ECU estimates system states respectively. 

Finally, ECUs transmit updated states to the edge coordinator which is able to design online scheduling or control strategies relying on its powerful computing capacity. With the issuance of new instructions, the edge computing supported ICPS enter the next sensing and control cycle.

As for the edge computing supported ICPS we will discuss, there are $n$ sensors and $m$ ECUs in the network. The communication topology of the edge computing layer is described as a static and undirected graph $\mathcal{G} = (\mathcal{V},\mathcal{E},\mathcal{H})$, where $\mathcal{V}$ is the set of ECUs, $\mathcal{E}$ is the set of links between ECUs, and $\mathcal{H}$ is the adjacency matrix. The set of ECU $i$-th's neighbors is $N_i=\{j|\forall(i,j)\in\mathcal{E}, i\neq j\}$.

\subsection{System Model}

Consider a linear discrete time-invariant system
\begin{equation}
x_{k+1} = Ax_k + w_k,   \label{eq:1}
\end{equation}
where the system matrix $A\in \mathbb{R}^{d\times d}$ is invertible. At time slot $k$, $x_k\in \mathbb{R}^d$ and $w_k\in \mathbb{R}^d$ are the system state and process noise, respectively. $w_k\sim \mathcal{N}(0,Q_k)$ is the Gaussian noise, and $x_0\sim \mathcal{N}(\Bar{x}_0,\Gamma_0)$ is the initial state.  

The observation of sensor $i$-th is as follows:
\begin{equation}
y_{i,k} = G_{i}x_k + v_{i,k},   \label{eq:2}
\end{equation}
where $G_{i}\in \mathbb{R}^{l\times d}$ is the observation matrix. At time slot $k$, $y_{i,k}\in \mathbb{R}^l$ and $v_{i,k}\in \mathbb{R}^l$ are the observation value and noise, respectively. $v_{i,k}\sim \mathcal{N}(0,U_{i,k})$ is the Gaussian noise. All $w_k$, $v_{i,k}$, and $x_0$ are independent from each other. 

For simplification, assume that each sensor observes a single dimension ($l=1$). Then the overall observation matrix and observation noise are obtained as
\begin{equation}
G = [G_{1}^\textrm{T},G_{2}^\textrm{T},...,G_{n}^\textrm{T}]^\textrm{T},   \label{eq:3}
\end{equation}
\begin{equation}
V_k = [v_{1,k}^\textrm{T},v_{2,k}^\textrm{T},...,v_{n,k}^\textrm{T}]^\textrm{T}.  \label{eq:4}
\end{equation}

In this paper, a sensor communicates its observation to an ECU over a channel with an additive white Gaussian noise (AWGN) using quadrature amplitude modulation (QAM). 

Define the reception variable as
\begin{equation}
  \gamma_{i,j,k}=\left\{\begin{aligned}
1, &\quad   \textrm{$y_{i,k}$ is received by ECU $j$-th at time slot $k$}\\
0, &\quad  \textrm{otherwise}.
\end{aligned}\right.
\label{eq:5}
\end{equation}

 Define the transmission power for the QAM symbol transmitted from sensor $i$-th to ECU $j$-th at time slot $k$ as $E_{i,j,k}$. The transmission success rate $\mu_{i,j,k}$ can be expressed as
\begin{equation}
\mu_{i,j,k} = \mathbb{P}[\gamma_{i,j,k}=1] = 1-\kappa^{E_{i,j,k}},
\label{eq:6}
\end{equation}
where $\kappa\triangleq \textrm{exp}(-\epsilon/(NW))\in(0,1)$, $\epsilon$ is a constant depending on the observation noise, $N$ is the AWGN noise power spectral density, and $W$ is the channel bandwidth\cite{z10}. 

% Regardless of the success of the transmission, the power $E_{i,j,k}$ is allocated to maintain a corresponding transmission success rate $\mu$, i.e. $\mathbb{E}[\gamma_{i,j,k}] = \mu_{i,j,k}$.

Define ECU $j$-th's observation and covariance of observation noise as:
\begin{equation}
 y_{j,k} = C_{j,k}Gx_k + C_{j,k}V_k,   \label{eq:7}
\end{equation}
\begin{equation}
R_{j,k} = C_{j,k}\textrm{diag}[U_{1,k},U_{2,k},...,U_{n,k}],  \label{eq:8}
\end{equation}
where $C_{j,k} \triangleq \textrm{diag}[\gamma_{1,j,k},...,\gamma_{n,j,k}]$ is the matrix for ECU $j$-th's observation reception at time slot $k$.

The following assumption is the basis for further analysis and stochastic sensing design. 

% \noindent\textbf{Assumption 1.} The spectral norms $||A||$, $||\Gamma_0||$, $||Q_k||$, and $||U_{i,k}||$ are upper bounded, denoting as $||A||\leq \bar{a}$, $||\Gamma_0||\leq \bar{p}$, $||Q_k||\leq \bar{q}$ and $||U_{i,k}||\leq \bar{u}$. 

\noindent\textbf{Assumption 1.}  $\{\gamma_{i,j,k}\}$ can be regarded as a set of binary variables with Bernoulli distributions, i.e. $\mathbb{E}[\gamma_{i,j,k}] = \mu_{i,j,k}$. $\gamma_{i,j,k}$ is independent from $\gamma_{r,s,t}$ when $i\neq r$, $j\neq s$ or $k\neq t$.

\noindent\textbf{Assumption 2.} The system is observable if each ECU collects all sensors' information at each time slot, i.e. ($A$,$G$) is observable.

As for edge sensing, each ECU adopts the distributed Kalman filtering algorithm \cite{z12}. After collecting information from linked sensors, ECU $j$-th's preprocessing result can be expressed as $\{ G^\textrm{T}C_{i,k}^\textrm{T}R^{-1}_{i,k}C_{i,k}G,G^\textrm{T}C_{i,k}^\textrm{T}R^{-1}_{i,k}y_{j,k}\}$. Then all ECUs share results with neighbors, and ECU $j$-th can obtain

\begin{equation}
S_{j,k} = \sum_{i\in N_j\cup\{j\}}G^\textrm{T}C_{i,k}^\textrm{T}R^{-1}_{i,k}C_{i,k}G,  \label{eq:9}
\end{equation}
\begin{equation}
y'_{j,k} = \sum_{i\in N_j\cup\{j\}}G^\textrm{T}C_{i,k}^\textrm{T}R^{-1}_{i,k}y_{j,k}. \label{eq:10}
\end{equation}

% where the information transmitted to neighbors by ECU $j$-th at time slot $k$ is $\{ G^\textrm{T}C_{i,k}^\textrm{T}R^{-1}_{i,k}C_{i,k}G,G^\textrm{T}C_{i,k}^\textrm{T}R^{-1}_{i,k}y_{j,k}\}$.

The recursions of ECU $j$-th's error covariance matrix $P_{j,k|k-1}$ and $P_{j,k|k}$ can be expressed as
\begin{align}
    P_{j,k|k-1} &= A P_{j,k-1|k-1}A^\textrm{T}+Q_k,\\
    P_{j,k|k} &= (P^{-1}_{j,k|k-1} + S_{j,k})^{-1}.
\end{align}

\subsection{Observability index}

In order to characterize the system's ability to achieve observability, the $L$-step observability and the observability index $\zeta$ are defined as  

\noindent\textbf{Definition 1.} The LTI system $(A,G)$ is $L$-step observable, if and only if 
\begin{equation}
    \textrm{rank}[G^\textrm{T}\quad A^\textrm{T}G^\textrm{T}\quad ...\quad (A^{L-1})^\textrm{T}G^\textrm{T} ] = d,\nonumber
\end{equation}
and $\zeta$ is defined as the observability index when $L = \zeta\leq d$ is the smallest integer that makes the condition hold \cite{z11}.

% \subsection{Distributed Edge Sensing}

%KF
%  ECUs adopt the distributed Kalman filtering algorithm \cite{z12} to estimate states of the system (1). Given ECU $j$-th's neighbors $N_j$, information collected from sensors and neighboring ECUs can be expressed as

% \begin{equation}
% S_{j,k} = \sum_{i\in N_j\cup\{j\}}G^\textrm{T}C_{i,k}^\textrm{T}R^{-1}_{i,k}C_{i,k}G,  \label{eq:9}
% \end{equation}
% \begin{equation}
% y'_{j,k} = \sum_{i\in N_j\cup\{j\}}G^\textrm{T}C_{i,k}^\textrm{T}R^{-1}_{i,k}y_{j,k}, \label{eq:10}
% \end{equation}
% where the information transmitted to neighbors by ECU $j$-th at time slot $k$ is $\{ G^\textrm{T}C_{i,k}^\textrm{T}R^{-1}_{i,k}C_{i,k}G_{k},G^\textrm{T}C_{i,k}^\textrm{T}R^{-1}_{i,k}y_{j,k}\}$.

% ECU $j$-th's prediction of the next step can be expressed as
% \begin{equation}
% P_{j,k|k-1} = A P_{j,k-1|k-1}A^\textrm{T}+Q_k,  \label{eq:13}
% \end{equation}
% \begin{equation}
% x_{j,k|k-1} = A x_{j,k-1|k-1}.  \label{eq:14}
% \end{equation}

% ECU $j$-th's estimation and error covariance matrix can be expressed as 
% \begin{equation}
% P_{j,k|k} = (P^{-1}_{j,k|k-1} + S_{j,k})^{-1},  \label{eq:11}
% \end{equation}
% \begin{equation}
% x_{j,k|k} = x_{j,k|k-1}+P_{j,k|k}[y'_{j,k}-S_{j,k}x_{j,k|k-1}].  \label{eq:12}
% \end{equation}

\subsection{Problem Formulation}

In this paper, we mainly focus on a stochastic scheduling problem over a considerable duration of time $T$. 

Define the cost function at time slot $k$ as 
\begin{equation}
    f_c(k) = \alpha\sum_{j =1}^m \textrm{tr}(P_{j,k|k}) + \beta\sum_{j =1}^m\sum_{i=1}^n E_{i,j,k}, 
\end{equation}
where $\alpha$ and $\beta$ are weights representing the preference in sensing accuracy and power consumption, respectively. $E_{i,j,k} = \textrm{ln}(1-\mu_{i,j,k})/\textrm{ln}(\kappa_{i,j,k})$ deduced from (6).

%  Define the sensing cost function $f_1(\mu_c)$ as
%  \begin{equation}
% f_1(\mu_{c}) = \sum_{k = 1}^T\sum_{j =1}^m \textrm{tr}(\mathbb{E}[P_{j,k|k}]),  \label{eq:15}
% \end{equation}
% where $\mu_c\in [0,1]^{mnT}$ is a vector related to $\{\mu_{i,j,k}|i\in\{1,2,...,n\}, j\in\{1,2,...,m\},k\in \{1,2,...,T\}\}$ containing all transmission success rate from each sensor to each ECU at overall time slots.

% Define the power consumption function $f_2(\mu_c)$ as 
% \begin{equation}
% f_2(\mu_c) = \sum_{k = 1}^T\sum_{j =1}^m\sum_{i=1}^n E_{i,j,k},  \label{eq:16}
% \end{equation}
% where $E_{i,j,k} = \textrm{ln}(1-\mu_{i,j,k})/\textrm{ln}(\kappa_{i,j,k})$ deduced from (6).

% 均匀感知——稳定性——更好的性能p0

To guarantee stability, referring to Theorem 2 in \cite{z8}, $\mathbb{E}[\textrm{tr}(P_{j,k|k})]$ is bounded, if the probability of ECU $j$-th's $L$-step observability $\phi^L_j> 1-\frac{1}{\bar{a}^{2L}}$, where $\bar{a}$ is the spectral norm of $A$. Although it is challenging to obtain exact observability probability, we can design functions to approximate it.

% Thus, it is beneficial to consider the probabilistic observability guarantee in stochastic sensor scheduling.

% To achieve uniform sensing performance over the duration $T$, we divide $T$ into $h$ small time intervals of length $L$, and analyze the relationship between observability probability and stability in each interval $L$. 

% Referring to Theorem 2 in \cite{z6}, we obtain that $\mathbb{E}[\textrm{tr}(P_{j,k|k})]$ is bounded for any time slot $k$ in $T$, if the requirement for observability probability $p_0\in (0,1)$, satisfying $p_0> 1-\frac{1}{\bar{a}^{2L}}$ for each interval.

\noindent\textbf{Problem 1.} Stochastic scheduling for edge sensing: 
\begin{align}
\label{yh}
    \mathop{\textrm{min}}_{\mu_{c}}&\mathop{\textrm{lim sup}}\limits_{T\rightarrow\infty} \frac{1}{T}\sum_{k=1}^T \mathbb{E}[f_c(k)],\\
     \textrm{s. t.}&\quad \mu_c\in [0,1]^{mnT}, \hat{\phi}_j^L(\mu_c) \geq p_0, \nonumber\\
     &\quad\forall j\in \{1,2,...,m\},\nonumber
\end{align}
where $\mu_c$ is a vector containing all transmission success rates $\mu_{i,j,k}$ from each sensor to each ECU at overall time slots. $\hat\phi_j^L(\mu_c)$ is the function approximating the observability probability $\phi_j^L(\mu_c)$, and $p_0$ is the probabilistic requirement for observability.

% In this problem, we need to design transmission power to maintain corresponding transmission success rate from sensors to ECUs. When these transmission success rates increase, the expected value of sensing cost will be lower and power consumption will be higher. Simultaneously, observability is easier to achieve. By solving this problem, we can make a balance between sensing accuracy and power efficiency with a basic observability requirement. 

\section{Deep Reinforcement Learning for Stochastic Sensor Scheduling}

In this section, we propose OIDM to solve Problem 1. The architecture of this method is shown in Fig. 2.

% The approximations of observability probability are provided by observability analysis, and DDPG is applied for optimization based on MDP formulation. Observability analysis also guides the design of action space in DDPG. 

\begin{figure}[htbp]
\centerline{\includegraphics[width=8cm]{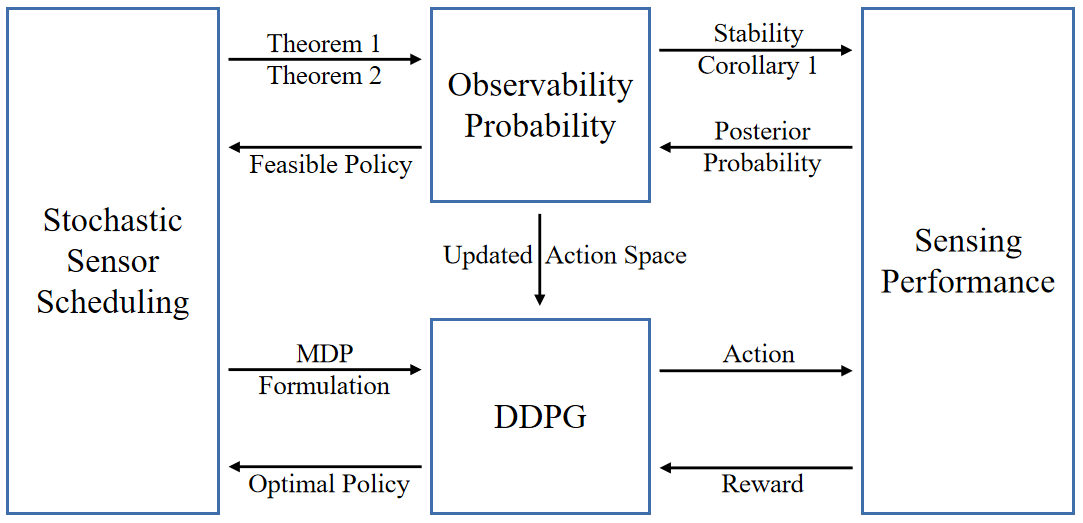}}
\caption{Basic logic flow of proposed OIDM.}
\label{fig:2}
\end{figure}

\subsection{Observability Analysis}

% Considering that the coupling of dimensions in the system matrix A is unfavorable for subsequent observability analysis, we transform the system equations into the Jordan canonical form. Since a non-singular transformation does not change the rank of the matrix, $(A,G)$ is observable if and only if $(J,\tilde{G})$. Furthermore, the observability index $\zeta$ of $(A,G)$ is equal to that of $(J,\tilde{G})$, which can be easily proven through observability criteria.

The Jordan decomposition is shown in (15). Since $A$ is invertible, there are no nilpotent blocks in $J$.
\begin{equation}
P^{-1}AP = J = \begin{bmatrix}
J_1 & & &\\
& J_2 & & \\
& & \ddots &\\
& & & J_\xi
\end{bmatrix}.
\end{equation}

Defining $\tilde{x}_k = P^{-1}x_k$  and $\tilde{G} = GP$, we obtain the Jordan form system equations (16) and (17). 
\begin{equation}
    \tilde{x}_{k+1} = J\tilde{x}_k + P^{-1}w_k,
\end{equation}
\begin{equation}
    y_{j,k} = C_{j,k}\tilde{G}\tilde{x}_k + C_{j,k}v_k. 
\end{equation}

$(A,G)$ is observable if and only if $(J,\tilde{G})$ is observable, and they share the same $\zeta$, which can be proven through observability criteria.
The new system is $L$-step observable if and only if the observability matrix $O_{j,L}$ is of column full rank. The observability matrix $O_{j,L}$ is defined as
\begin{equation}
 O_{j,L} = \left[
 \begin{array}{ccc}
     (\Theta_j\otimes I_n)C_{k_0}\tilde{G}\\
  (\Theta_j\otimes I_n)C_{k_0+1}\tilde{G}J\\
  (\Theta_j\otimes I_n)C_{k_0+2}\tilde{G}J^2\\
  ...\\
  (\Theta_j\otimes I_n)C_{k_0+L-1} \tilde{G}J^{L-1}
 \end{array}
 \right] ,\label{eq:20}
\end{equation}
where $ \Theta_j= \textrm{diag}[D_{(j,:)}]$, $D = I_m + H$, $C_k = [C_{1,k};...;C_{m,k}]$, $k_0$ is the initial time slot, and $L\geq \zeta$.

Since it is difficult to analyze the observability condition $O_{j,L}$ of full rank in a stochastic sense, we derive novel linear approximations based on $\zeta$ in Theorem 1 and Theorem 2. 

% \noindent\textbf{Theorem 2.}  Linear approximations of observability criteria based on $\zeta$ are provided as 

\noindent\textbf{Theorem 1.} The necessary condition to guarantee the $L$-step observability is
\begin{equation}
\forall j,\quad \sum_{k=1}^L\sum_{i\in N_j\cup \{j\}}\textrm{tr}(C_{i,k})\geq (\zeta-1)\textrm{rank}(\tilde{G}) + 1. \nonumber
\end{equation} 
\noindent\textit{Proof:} Considering the system observability through ECU $j$-th, we obtain
\begin{align}
    \textrm{rank}(O_{j,L}) &=  \textrm{rank}(\Big[[I_d,J^\textrm{T},...,(J^{L-1})^\textrm{T}]\nonumber\\&\quad\quad\quad\quad\quad\quad(I_L\otimes (\tilde{G}^\textrm{T}C_k^\textrm{T}(\Theta_j\otimes I_n)^\textrm{T}   ))\Big]^\textrm{T})  \nonumber\\
    &\leq \min \Big\{ d, \sum_{k=1}^L \textrm{rank}(I_L\otimes (\tilde{G}^\textrm{T}C_k^\textrm{T}(\Theta_j\otimes I_n)^\textrm{T}   )) \Big\} \nonumber\\
    &\leq \sum_{k=1}^L\sum_{i\in N_j\cup\{j\}}\min\Big\{\textrm{rank}(\tilde{G}^\textrm{T}), \textrm{rank}(C_{i,k}^\textrm{T})  \Big\}\nonumber\\
    &\leq \sum_{k=1}^L\sum_{i\in N_j\cup\{j\}} \textrm{tr}(C_{i,k}). \nonumber
\end{align}

Because $\zeta$ is the smallest integer that makes the observability condition holds, we have $\zeta\cdot\textrm{rank}(\tilde{G})\geq d \geq (\zeta-1) \textrm{rank}(\tilde{G})+1$. When the system is observable, $\textrm{rank}(O_{j,L}) = d$. Thus, based on the equality and inequalities above, we obtain $\sum\limits_{k=1}^L\sum\limits_{i\in N_j\cup \{j\}}\textrm{tr}(C_{i,k}) \geq (\zeta-1)\textrm{rank}(\tilde{G}) + 1.$ $\hfill\blacksquare$

% Due to the lower bound of $\zeta$ shown in Theorem 1, we obtain
% \begin{equation}
%     \zeta\cdot\textrm{rank}(\tilde{G})\geq d \geq (\zeta-1) \textrm{rank}(\tilde{G})+1. \nonumber
% \end{equation}

% When the system is observable, based on inequalities above, the following inequality for one ECU is obtained.

% \begin{equation}
% ~~~~~~~~~\sum_{k=1}^L\sum_{i\in N_j\cup \{j\}}\textrm{tr}(C_{i,k})\geq (\zeta-1)\textrm{rank}(\tilde{G}) + 1. ~~~~~~~\hfill\blacksquare\nonumber
% \end{equation}

% \noindent\textit{Proof:} After Jordan transformation, the system possesses new properties. We utilize these properties to prove this theorem.

\noindent\textbf{Theorem 2.} The sufficient condition to guarantee the $L$-step observability is 
\begin{equation}
\exists\mathcal{K}, \forall k\in\mathcal{K}, \forall j, \forall i\in \mathcal{M}, \sum_{\tau\in N_j\cup\{j\}} C_{\tau,k}[i,i]\geq 1, \nonumber
\end{equation}
where $\mathcal{K}$ denotes the index set of any segment of length not less than $\zeta$ within the interval of $L$, and $\mathcal{M}$ denotes the index set of rows in $\tilde{G}$, which can observe the first dimension of any Jordan block in $J$.

\noindent\textit{Proof:} As shown in Fig. 3, due to the structure of Jordan form, in a Jordan block, a dimension can be deduced only by observations of dimensions above it within $\zeta$ steps. Based on Assumption 2, for each Jordan block, there is at least one row in $\tilde{G}$ observing its first dimension. Thus, it is possible that $O_{j,L}$ can be of full rank when ECU $j$-th and its neighbors only observe each first dimension of all Jordan blocks in $J$ during the interval $L$. Furthermore, if ECU $j$-th and its neighbors observe all rows in $\mathcal{M}$ at every time slot $k$ (more than $\zeta$ times) in any $\mathcal{K}$, $O_{j,L}$ is of full rank. This method of proof is similar to that used in \cite{z13}. 
$\hfill\blacksquare$

\begin{figure}[htbp]
\centerline{\includegraphics[width=5cm]{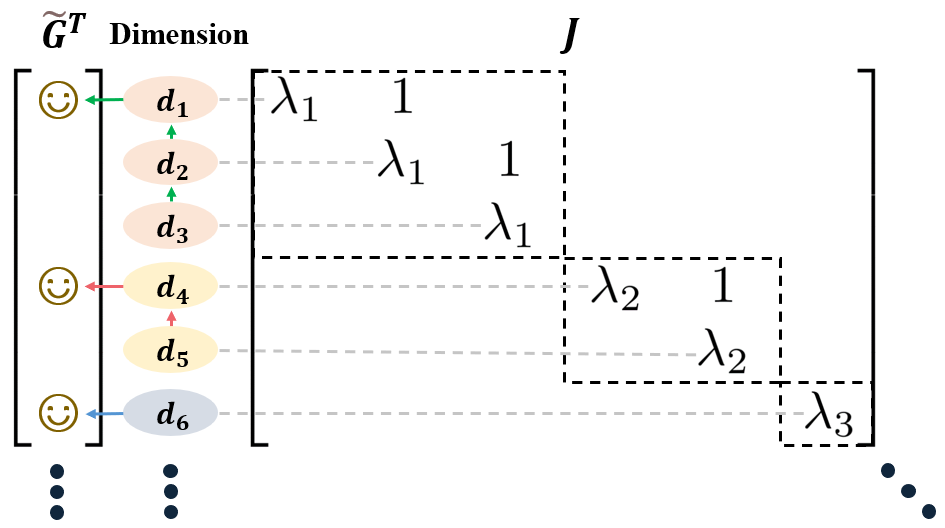}}
\caption{An example of the topology of $(J,\tilde{G})$.}
\label{fig:3}
\end{figure}
% The lower bound and the upper bound of observability probability can be derived from the sufficient condition and the necessary condition in Theorem 2, respectively. These boundary functions can serve as $\hat{\phi}_j^L(\mu_c)$ to approximate the observability probability $\phi^L_j(\mu_c)$.  

% $\phi_j(\mu_c)$ is certainly not less than the probability of satisfying the sufficient condition and not greater than the probability of satisfying the necessary condition. 

\noindent\textbf{Corollary 1.}  In each interval of length $L$, the ECU $j$-th's observability probability  $\phi^L_j(\mu_c)$ is bounded as
\begin{equation}
    \underline\varphi_j \leq \phi^L_j(\mu_c) \leq \overline{\varphi}_j,\nonumber
\end{equation}
where $\hat\mu_{i,j,k} = 1-\prod\limits_{\tau\in N_j\cup\{j\}}(1-\mu_{i,\tau,k})$, $\varpi = (|N_j|+1)nL$, $\varrho = (\zeta-1)\textrm{rank}(\tilde{G}) + 1$, $\mathcal{K}_\iota^\eta$ is the $\eta$-th index set of time periods of length $\iota$ within the interval of $L$, $\Psi_{\tau}^\eta$ is the $\eta$-th index set of successful sensor transmissions, in which $\varpi$ transmissions has $\tau$ successful transmissions, 
$$\underline\varphi_j = \sum_{\iota=\zeta}^{L} \sum_{\eta=1}^{C_{L}^{\iota}} \prod_{\tau\in\mathcal{M}} [\prod_{k\in\mathcal{K}_\iota^\eta}\hat\mu_{\tau,j,k} \prod_{k\in\bar{\mathcal{K}}_\iota^\eta}(1-\hat\mu_{\tau,j,k})],$$
$$\overline\varphi_j = \sum_{\tau=\varrho}^{\varpi}\sum_{\eta=1}^{C_{\varpi}^{\tau}}[\prod_{(i,k)\in\Psi_{\tau}^{\eta}}\mu_{i,j,k}\prod_{(i,k)\in\bar\Psi_{\tau}^\eta}(1-\mu_{i,j,k})].$$

% 与自己对比：分析概率，得到确定保障
% 与师兄对比：不需要对观测矩阵限制，利用Jordan分解和observability index，得到更广义的结果

% zeta桥梁
% Different from observability analysis in deterministic scheduling \cite{z6}, we bound the observability probability by analyzing the number of successful sensor transmissions (Corollary 1). Furthermore, based on these bounds, a feasible policy guaranteeing probabilistic observability can be obtained for further seeking the optimal one. Compared with \cite{z8}, we provide linear approximations of observability criteria with multiple ECU cooperation (Theorem 2), and guide the design of $L$ based on observability index $\zeta$ (Theorem 1). Additionally, we utilize Jordan transformation to overcome the binary constraint imposed on the observation matrix, and obtain more general results.  

\subsection{MDP Formulation}

 Due to the Markov property of the sensing cost $P_{j,k|k}$, Problem 1 can be formulated into an MDP denoted as $(\mathcal{S},\mathcal{A},\mathcal{P},\mathcal{C})$. Details are described as follows.

 1) The state space $\mathcal{S} \triangleq \mathbb{R}^{m\times d\times d}$ includes all ECUs' estimation error covariances, and the state at time slot $k$ can be indicated by $s_k \triangleq (P_{1,k|k},...,P_{m,k|k})$.
 
 %$s_k \triangleq (P_{1,k|k},...,P_{m,k|k})/P_0$, where $P_0$ is the initial estimation error covariance and dividing by $P_0$ is used for normalization. 

 2) The action space $\mathcal{A}\triangleq [0,1]^{m\times n}$ consists of each transmission success rate, and the action at time slot $k$ can be indicated by $a_k \triangleq (\mu_{1,1,k},...,\mu_{n,m,k})$.

 3) The transmission probability $\mathcal{P}(s'|s,a):\mathcal{S}\times\mathcal{A}\times\mathcal{S}\rightarrow \mathbb{R}$ can be defined as $\mathcal{P}(s_{k+1}|s_k,a_k)\triangleq \prod\limits_{(i,j)\in\Omega_k}\mu_{i,j,k}\prod\limits_{(i,j)\in\bar\Omega_k}(1-\mu_{i,j,k})$, where $\Omega_k = \{(i,j)|\gamma_{i,j,k}=1\}$ inferred by $(s_{k+1},s_k)$.
 
% s->s'决定选或不选，a决定概率的大小

 4) The cost function $\mathcal{C}(s,a): \mathcal{S}\times \mathcal{A}\rightarrow \mathbb{R}$ involves the sensing cost and power consumption, and the cost produced at time slot $k$ can be indicated by $\mathcal{C}(s_k,a_k)\triangleq -\sum\limits_{j =1}^m \Big[\alpha\textrm{tr}(P_{j,k|k}) + \beta\sum\limits_{i=1}^n E_{i,j,k}\Big]$, where $E_{i,j,k}$ can be deduced by $a_k$ according to (6). 

\subsection{OIDM}
%介绍DDPG方法

% DDPG is developed to optimize the stochastic scheduling \cite{z9}. DDPG is a model-free reinforcement learning algorithm that combines the advantages of deep learning and deterministic policy gradient methods. 

DDPG utilizes an actor-critic framework, where an actor network $\pi(s|\theta_a)$ with parameters $\theta_a$ learns the optimal policy and a critic network $Q(s,a|\theta_c)$ with parameters $\theta_c$ estimates the value function. The actor network directly maps states to actions, while the critic network evaluates the quality of the actor's actions. The proposed OIDM combines DDPG with observability analysis as shown in Algorithm 1.

% 从条件到转化为指导动作空间
% 1) First, considering that the power consumption $E\rightarrow +\infty$ as the transmission success rate $\mu\rightarrow 1$ (6), it is impractical to expend such a high amount of power to maintain an extremely high success rate. There is redundancy in the action space $\mathcal{A}= [0,1]^{m\times n}$, which hampers the network's ability to efficiently search for the optimal policy. 

%动作空间优化
1) First, after selecting an $L$ as the desired scale for guaranteeing observability, a feasible solution can be obtain from the $L$-step observability analysis mentioned above to guide the design of the action space. Let $\underline\varphi_j(\mu_c) = p_0$ with identical elements in $\mu_c$ for a certain $p_0$ satisfying Corollary 1. This solution is denoted as $\bar\mu_c^j = (\bar\mu_{fea}^j,...,\bar\mu_{fea}^j)_{m\times n\times T}$. Due to power consumption in the objective function, there exists solutions with better sensing performance than this feasible one in $[0,\bar\mu_{fea}^j]$. In this way, define $\bar\mu_{fea} = \textrm{max}\{\bar\mu_{fea}^j|j\in\{1,...,m\}\}$. Similarly, we can obtain a lower bound $\underline{\mu}_{fea}$ from $\bar\varphi_j(\mu_c) = p_0$. Denote the solution as $\underline{\mu}_c^j = (\underline{\mu}_{fea}^j,...,\underline{\mu}_{fea}^j)_{m\times n\times T}$. Due to higher transmission success rate, there exists solutions better guaranteeing observability in $[\underline{\mu}_{fea}^j,1]$. Thus define $\underline{\mu}_{fea} = \textrm{max}\{\underline{\mu}_{fea}^j|j\in\{1,...,m\}\}$, and update the action space as $\mathcal{A}'\triangleq[\underline{\mu}_{fea},\bar\mu_{fea}]^{m\times n}$, guaranteeing $\hat{\phi}_j^L(\mu_c) = \bar\varphi_j(\mu_c)\geq p_0$. We can further search for the optimal policy based on the feasible one in this new action space $\mathcal{A}'$, enabling a balance between the accuracy and efficiency with a basic observability guarantee.

% 解释DDPG中关键公式，方便伪代码引用
2) Then DDPG is adopted to seek the optimal policy in the updated action space $\mathcal{A}'$. Replay Buffer is also utilized to improve the efficiency of learning. The action network $\pi(s|\theta_a)$ is updated by gradient ascent as
\begin{equation}
    \nabla_{\theta_a}\mathcal{J}\approx \mathbb{E}[\nabla_a Q(s,a|\theta_c)|_{s=s_k,a=\pi(s_k)}\nabla_{\theta_a}\pi(s|\theta_a)|_{s_k}].
\end{equation}

The critic network $Q(s,a|\theta_c)$ is updated by minimizing the loss function as
\begin{equation}
    \mathcal{L}_k = \mathbb{E}_{(s,a,c,s')}[(Q(s_k,a_k|\theta_c)- r_k)^2],
\end{equation}
where $r_k = \mathcal{C}(s_k,a_k)+\vartheta Q(s_{k+1},\pi(s_{k+1}|\theta_a)|\theta_c)$.

In order to improve the convergence of the algorithm, Soft Target Update is employed as
\begin{align}
    \theta_{c}^{targ}\leftarrow \varrho\theta_c + (1-\varrho)\theta_{c}^{targ},\nonumber\\
    \theta_a^{targ}\leftarrow \varrho\theta_a+(1-\varrho)\theta_a^{targ},
\end{align}
 where $\varrho\in (0,1)$ is the update rate.

The justification for approximating the solution to Problem 1 with a discount factor $\vartheta\in(0,1)$ has been given in \cite{z7}.

 \begin{algorithm}[htbp]
\caption{OIDM} 
\hspace*{0.02in} {\bf Input:} 
Initialization parameter $\theta_{c}^0$, $\theta_a^0$, $\Gamma_0$; Buffer capacity $B$, Mini-batch size $b$; Episode $M$, Period $T$; Update rate $\varrho$; Probabilistic requirement for observability $p_0$; Weights $\alpha$, $\beta$.\\
\hspace*{0.02in} {\bf Output:} 
Sensing policy $\pi^*(s|\theta_a)$
\begin{algorithmic}[1]
\State Initialize the critic $Q(s,a|\theta_c)$, the actor $\pi(s|\theta_a)$ and their target networks with $\theta_{c}^0$ and $\theta_a^0$.
\State Select an $L$ as the desired scale for guaranteeing observability with a probability of $p_0$.
\State Update the action space as $\mathcal{A}'\triangleq[\underline\mu_{fea},\bar\mu_{fea}]^{m\times n}$ based on observability analysis.
\State Initialize the replay buffer $\mathcal{B}$ with capacity $B$.
\For{episode=$1:M$}
    \State Initialize the initial state $s_0$ with $\Gamma_0$.
    \For{$k=1:T$}
        \State Obtain the transmission success rate $a_k$ from $\pi(s_k|\theta_a)$ and add a Gaussian noise. 
        \State Compute $\{s_k,a_k,c_k,s_{k+1}\}$ according to (13).
        \State Put the transition $\{s_k,a_k,c_k,s_{k+1}\}$ into $\mathcal{B}$.
        \State Randomly pick a mini-batch of $b$ transitions $\{s_i,a_i,c_i,s_{i+1}\}$ from $\mathcal{B}$.
        \State Update the critic network according to (20).
        \State Update the actor network according to (19).
        \State Update the target networks according to (21).
    \EndFor
\EndFor
\end{algorithmic}
\end{algorithm}

% gailv duocishengcheng 概率多生成的思想来解决无法直接分析随机矩阵的秩

\section{Application to Hot Rolling Process}
In this section, OIDM is applied to estimate slab temperature in industrial hot rolling process as shown in Fig. 4.

\begin{figure}[htbp]
\centerline{\includegraphics[width=8.5cm]{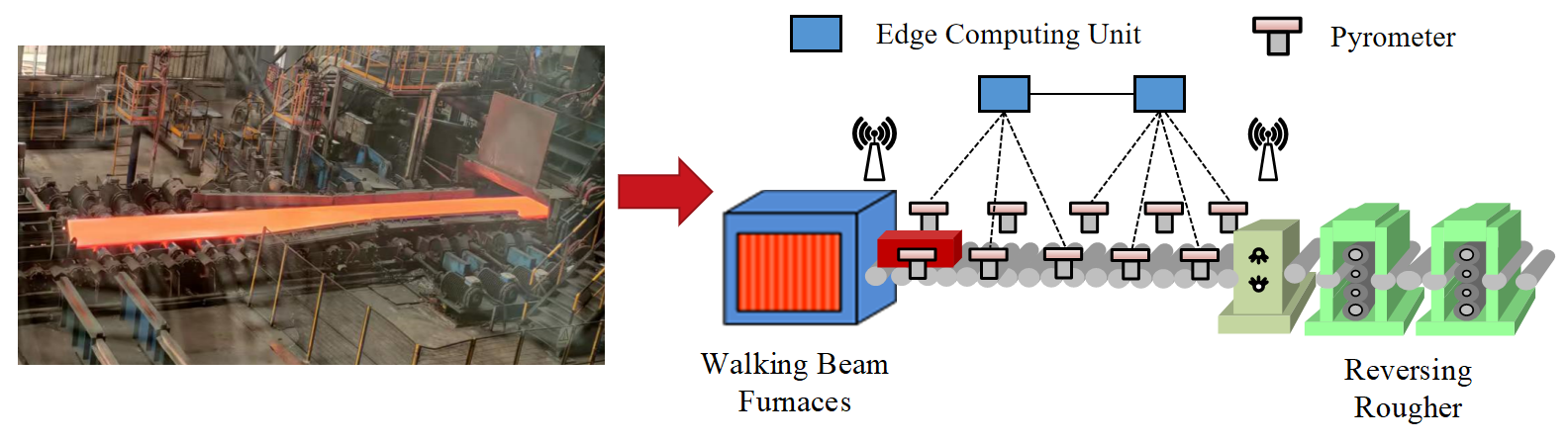}}
\caption{One typical scenario of ICPS: temperature estimation in industrial hot rolling process.}
\label{fig:4}
\end{figure}

\subsection{Physical Model}

We are dedicated to estimating the slab temperature at the entry of the rougher. In this industrial scenario, air cooling and heat conduction are the primary factors influencing the variation of slab temperature. Neglecting the transverse temperature change of the slab, we build a discrete model referring to \cite{z14} and \cite{z15}. 
To facilitate computational simplification, the slab can be divided into $\tau$ sections. For the $s$-th section $x^s$, we divide the length and thickness into $\tau_s$ and $\nu$ lattices, respectively. 

% As shown in Fig. \ref{fig:5}, each section is treated similarly. 

% \begin{figure}[htbp]
% \centerline{\includegraphics[width=8.7cm]{5.PNG}}
% \caption{The division of the slab.}
% \label{fig:5}
% \end{figure}

Define $x^s$ = $[(x^{s(1)})^\textrm{T},(x^{s(2)})^\textrm{T},...,(x^{s(\tau_s)})^\textrm{T}]^\textrm{T}$, $
x^{s(j)} = [x^s_{1,j},x^s_{2,j},...,x^s_{\nu,j}]^\textrm{T}$, $\forall j \in \{1,2,...,\tau_s\}$. Then the temperature of the $s$-th section can be modeled as
\begin{equation}
    x^s(k+1) = Fx^s(k)-u(k)+w(k), 
\end{equation}
with
\begin{equation}
\setlength{\arraycolsep}{1.2pt}
\scriptsize{
\mathbb{Q} = \left[
 \begin{array}{ccccc}
     -2 & 2 & 0 & \dots & 0\\
1 & -2 & 1 & \ddots & \vdots\\
0&\ddots&\ddots&\ddots&0\\
\vdots&\ddots&1&-2&1\\
0&\dots&0&2&-2
 \end{array}
 \right], \Lambda = \left[
 \begin{array}{cccc}
     \omega I_{\nu} & 0 & \dots & 0 \\
-\omega I_{\nu} & \omega I_{\nu} &  & \vdots \\
\vdots&\ddots&\ddots&0\\
0&\dots&-\omega I_{\nu} & \omega I_{\nu}
 \end{array}
 \right],\nonumber}
\end{equation}

% \begin{equation}
%     \tiny{\Lambda = \left[
%  \begin{array}{cccc}
%      \omega I_{\nu} & 0 & \dots & 0 \\
% -\omega I_{\nu} & \omega I_{\nu} &  & \vdots \\
% \vdots&\ddots&\ddots&0\\
% 0&\dots&-\omega I_{\nu} & \omega I_{\nu}
%  \end{array}
%  \right], \nonumber}
% \end{equation}

\begin{equation}
    \scriptsize{\mathbb{P}_j = \left[
    \begin{array}{c}
    \beta(x^s_{1,j})((x^s_{1,j}(k))^4-x^4_\infty)\\
\textbf{0}\\
\beta(x^s_{v,j})((x^s_{v,j}(k))^4-x^4_\infty)
 \end{array}
\right],\nonumber}
\end{equation}
where $F$ = $\Lambda+\textrm{diag}\{D_1\mathbb{Q},...,D_{\tau_s}\mathbb{Q}\}$, $u(k) = [\mathbb{P}_1^\textrm{T},...,\mathbb{P}^\textrm{T}_{\tau_s}]^\textrm{T}$, 
$D_j$ = $\textrm{diag}\{\alpha(x^s_{1,j}(k)),...,\alpha(x^s_{\nu,j}(k))\}$, $\alpha(x^s_{i,j}(k)) = \Delta t\lambda^s_{i,j}/(\Delta b^2\rho^s_{i,j}c^s_{i,j})$, $\beta(x^s_{p_*,j}) = 2\Delta t\sigma_0\varepsilon/(\Delta b\rho^s_{p_*,j}c^s_{p_*,j})$, $\omega = \upsilon/(2\Delta l) $, $i = 1,...,\nu$, $j = 1,2,...,\tau_s$, $p_* = 1, \nu$.

\subsection{Parameter Selection}

% 小段数量和长度还没定，要在后面讨论
For simplification, the differences in slab density, specific heat capacity, thermal conductivity, thermal radiation coefficient and initial temperature can be neglected. Additionally, we assume that $\mathbb{P}_j$ is a constant matrix, because the slab temperature experiences minimal change within a brief period (e.g. 0.2 seconds). Thus, the slab cools down at a constant rate $u(k)$ within $T$. Finally, a discrete-time linear time-invariant system is obtained. 

% 热轧 G、kappa要重新改一下
The values of hot-rolling parameters are as follows: specific heat capacity $c$ = 460 J/(kg·K), density $\rho$ = $7.9\times 10^3$ $\mathop{\textrm{kg/m}}^3$, thermal conductivity $\lambda$ = 40 W/(m·K), thermal radiation coefficient $\varepsilon$ = 0.85, Stefan Boltzmann constant $\sigma_0$ = $5.67\times 10^{-8}$ W/($\mathop{\textrm{m}}^2$·$\mathop{\textrm{K}}^4$), environment temperature $x_\infty$ = 325K, initial temperature of slab $x_0$ = 1180K, the number of lattices in length $\tau_s$ =  10, the number of lattices in thickness $\nu$ =  3, thickness discretization step $\Delta b$ =  0.01 m, length discretization step $\Delta l$ =  5 m, conveyor speed $\upsilon$ = 5m/s, time step $\Delta t$ =  0.2 s, and total time slots $T$ = 1000. 

% 网络 alpha\beta
We deploy $m$ = 2 ECUs and $n$ = 10 sensors above the slab, and ECUs are fully connected to each other. The values of network parameters are as follows: $Q = 0.1I_d, R = 0.01I_n, P_{j,0|0} = I_d$, $G_{i,k} = \Big[[G_{i,k}]_{3i-2} = 1, \textrm{else}\quad0 \Big]$, $p_0 = 0.95$, $L = 10$, and $\alpha = \beta = 0.1$. $\kappa = 0.3$ for ECU 1st; 0.4 for ECU 2nd.  If ECU $j$-th cannot link sensor $i$-th in actual industrial scenarios, we can set $\kappa_{i,j}\rightarrow 1$ correspondingly. 

%训练
We use one hidden layer with 1024 nodes in the actor and critic neural networks, respectively. The learning rates $lr_a$ and $lr_c$ are $1e^{-4}$ and $1e^{-3}$ respectively. The Adam optimizer is adopted. The values of training parameters are as follows: the number of episodes $M = 200$, the number of steps in each episode $T = 1000$, size of reply buffer $B = 10^5$, mini-batch size $b = 64$, variance of exploration noise $\sigma = 0.02$, discount factor $\vartheta = 0.99$, and soft update factor $\varrho = 0.001$.

\subsection{Results and Discussion}

% 收敛+训练时间可行性分析
% Based on the above parameters, the training process of OIDM is shown in Fig. 5. After the buffer is filled to capacity, it takes around 50 episodes for the cost to converge. We trained for 200 episodes on our 2.6 GHz Intel Core i7 processor, with each episode taking no longer than 25 seconds. 

% \begin{figure}[htbp]
% \centerline{\includegraphics[width=7cm]{44.png}}
% \caption{A training process of OIDM.}
% \label{fig:6}
% \end{figure}

% 总代价和感知代价的对比

We compare sensing performance of OIDM with other sensing methods as shown in Fig. 5. We repeated each experiment for $H=100$ times, and used the average of the results to approximate the expected cost. The sensing prioritized method (SPM) always selects the maximum action in $\mathcal{A}'$. The random scheduling method (RSM) selects an action in $\mathcal{A}'$ randomly. The periodic scheduling method (PSM) selects the maximum and minimum action in $\mathcal{A}'$ alternately. When $\beta \rightarrow 0$, sensing cost is the determining factor, and the performance of SPM and OIDM is nearly identical. As $\beta$ increases, power consumption is gradually prioritized, and SPM causes the maximum cost among these methods. PSM and RSM are largely unaffected by $\beta$, and they maintain a uniform but mediocre performance. Our OIDM consistently achieves a good balance between accuracy and efficiency.

\begin{figure}[htbp]
\centering  
\subfigure[$\beta = 10\alpha$]{
\label{Fig.sub.4}
\includegraphics[height=1.84cm]{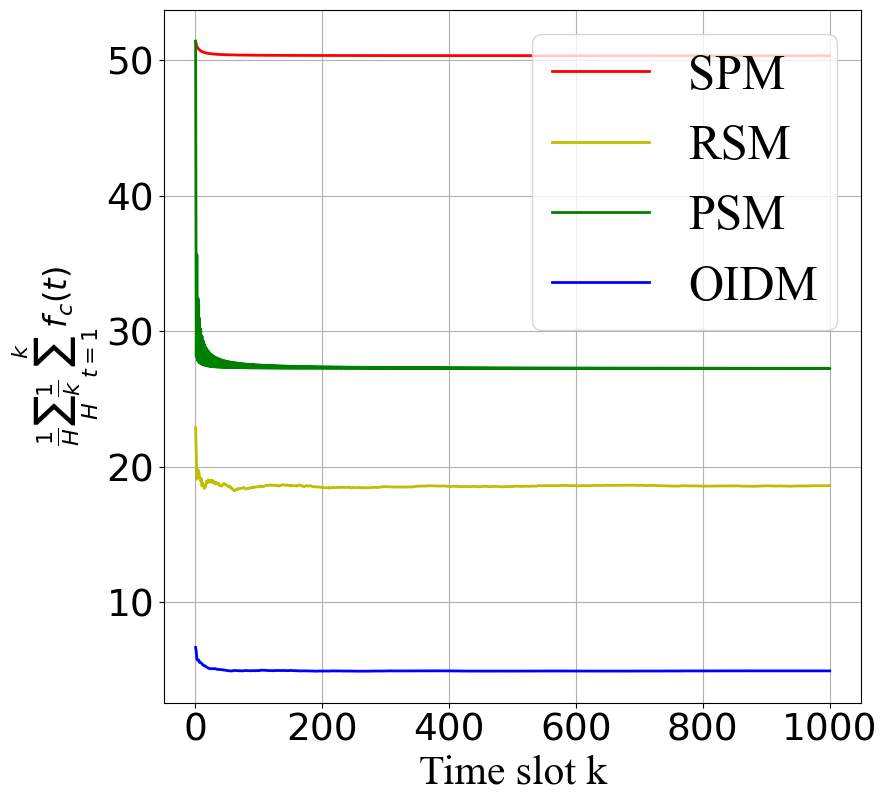}}\subfigure[$\beta = \alpha$]{
\label{Fig.sub.1}
\includegraphics[height=1.84cm]{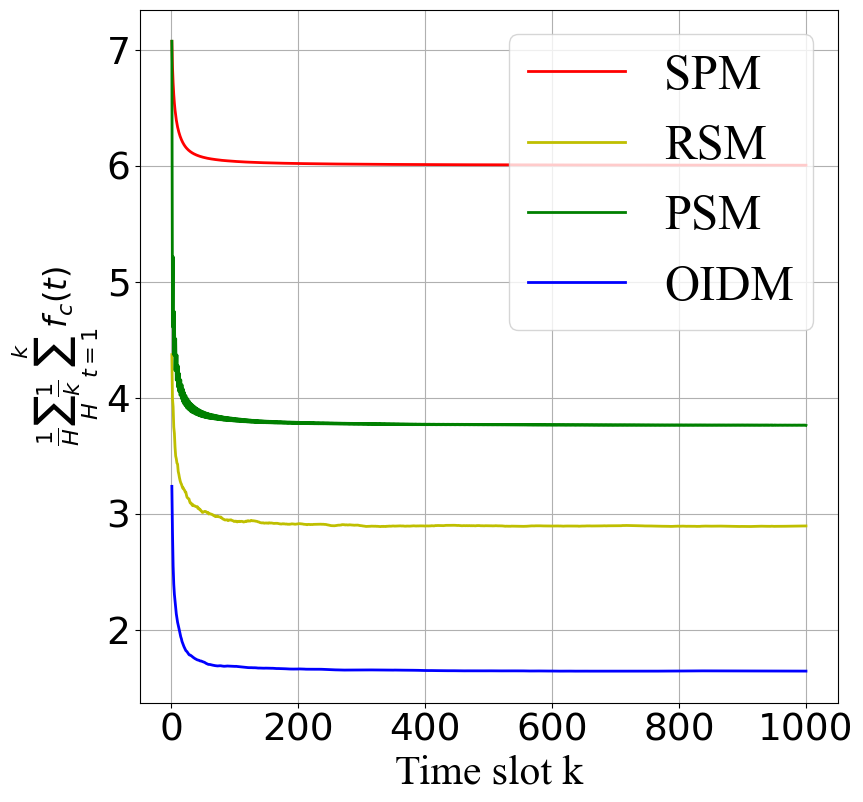}}\subfigure[$\beta = 0.1\alpha$]{
\label{Fig.sub.2}
\includegraphics[height=1.84cm]{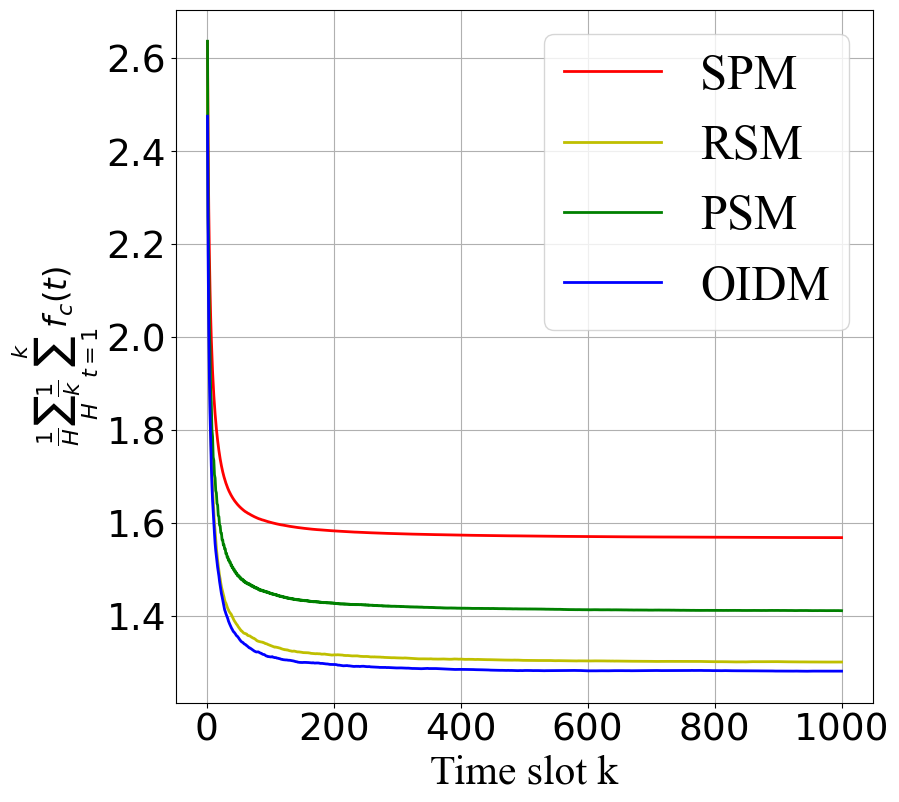}}\subfigure[$\beta = 0$]{
\label{Fig.sub.3}
\includegraphics[height=1.84cm]{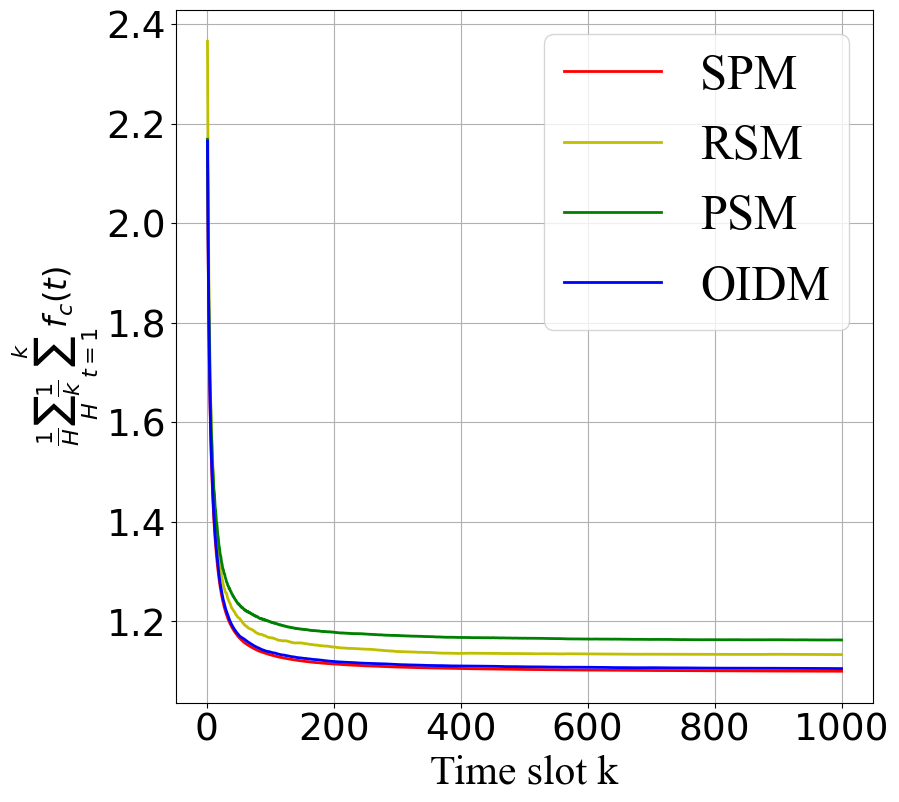}}
\caption{Comparisons of average cost with different $\beta$.}
\label{fig:7}
\end{figure}

% \begin{table}[htbp]
% \caption{Observability Probability Comparison with Different $(L,\beta)$}
% \begin{center}
% \begin{tabular}{c|c|c|c|c}
% \hline\hline
% \textbf{OIDM}&\textbf{$\beta$ = 0.01$\alpha$}&\textbf{$\beta$ = 0.1$\alpha$}&\textbf{$\beta$ = 1$\alpha$}&\textbf{$\beta$ = 10$\alpha$} \\
% \hline
% \textbf{$L$ = 5}& 1  & 1 & 0.9380 & 0.9223    \\

% \textbf{$L$ = 10}& 1 & 0.9987 & 0.4037 &0.3063  \\
% \hline\hline
% \end{tabular}
% \label{beta}
% \end{center}
% \end{table}

% 可观性概率上下界平均, L长度的影响

Then we concentrate on the design of $L$ based on guaranteeing observability. The observability probabilities with different $(L,\beta)$ are compared in Table \uppercase\expandafter{\romannumeral 1}. The 10-step observability is regarded as the standard. We checked the observability of each 10-step interval in $T$, and repeated each experiment for $H=100$ times. The average of posterior observability probabilities were obtained. We can notice that a smaller $L$ provides a stronger guarantee of observability. As $\beta$ increases, the gap becomes noticeable, but there exists a basic guarantee of probabilistic observability. 

% A relatively small $L$ would be better to guarantee the effectiveness of our approximation when $\beta$ is large. 

\begin{table}[htbp]
\caption{Observability Probability Comparison with Different $(L,\beta)$}
\begin{center}
\begin{tabular}{c|c|c|c|c|c}
\hline\hline
\textbf{OIDM}&\textbf{$\beta$ = 0}&\textbf{$\beta$ = 0.1$\alpha$}&\textbf{$\beta$ = 1$\alpha$}&\textbf{$\beta$ = 10$\alpha$}&\textbf{$\beta$ = 100$\alpha$} \\
\hline
\textbf{$L$ = 5}& 1  & 1 & 0.9380 & 0.9223 & 0.9223   \\

\textbf{$L$ = 10}& 1 & 0.9987 & 0.4037 &0.3063 & 0.3063  \\
\hline\hline
\end{tabular}
\label{beta}
\end{center}
\end{table}

Finally, we focus on the action space of OIDM with different $(L,p_0)$. As shown in Fig. 6, a smaller $L$ provides a more compressed action space, possibly leading to the loss of policies with better sensing performance. Conversely, when $L$ is excessively large, there will be negligible compression of the action space, limiting the effectiveness of this method. For a fixed $L$, before the probabilistic requirement $p_0$ approximates $1$, the action space expands as $p_0$ increases. However, the action space turns to $0$ when $p_0=1$.

% Thus, we need to select an appropriate $L$ to make a trade-off. 
 
\begin{figure}[htbp]
\centering  
\subfigure[Upper bound]{
\label{Fig.sub.1}
\includegraphics[height=2.5cm]{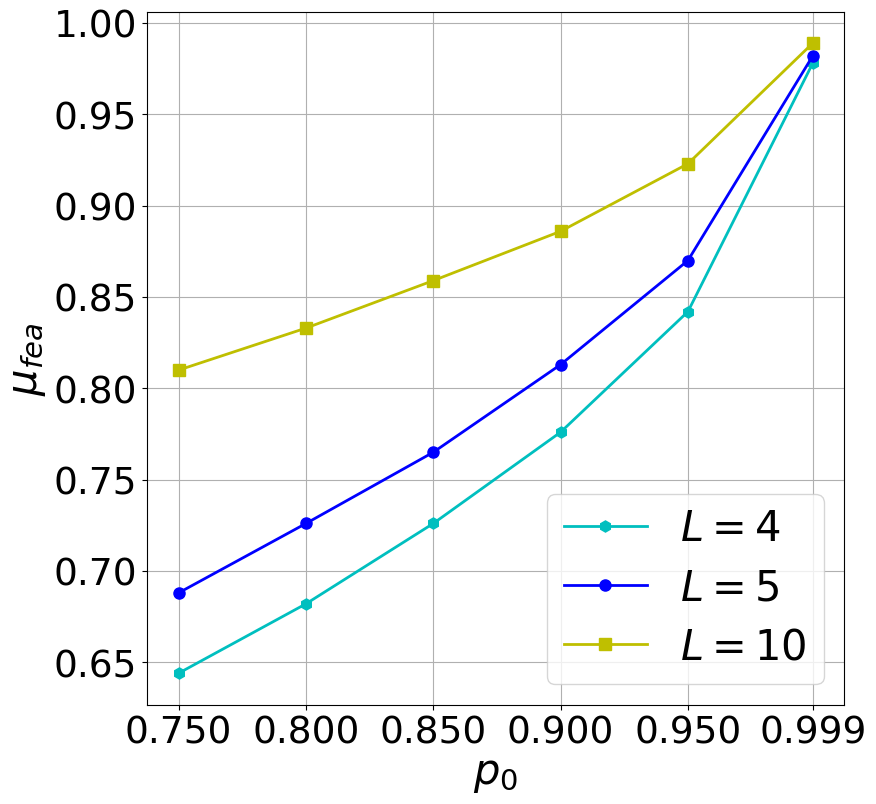}}\subfigure[Lower bound]{
\label{Fig.sub.2}
\includegraphics[height=2.5cm]{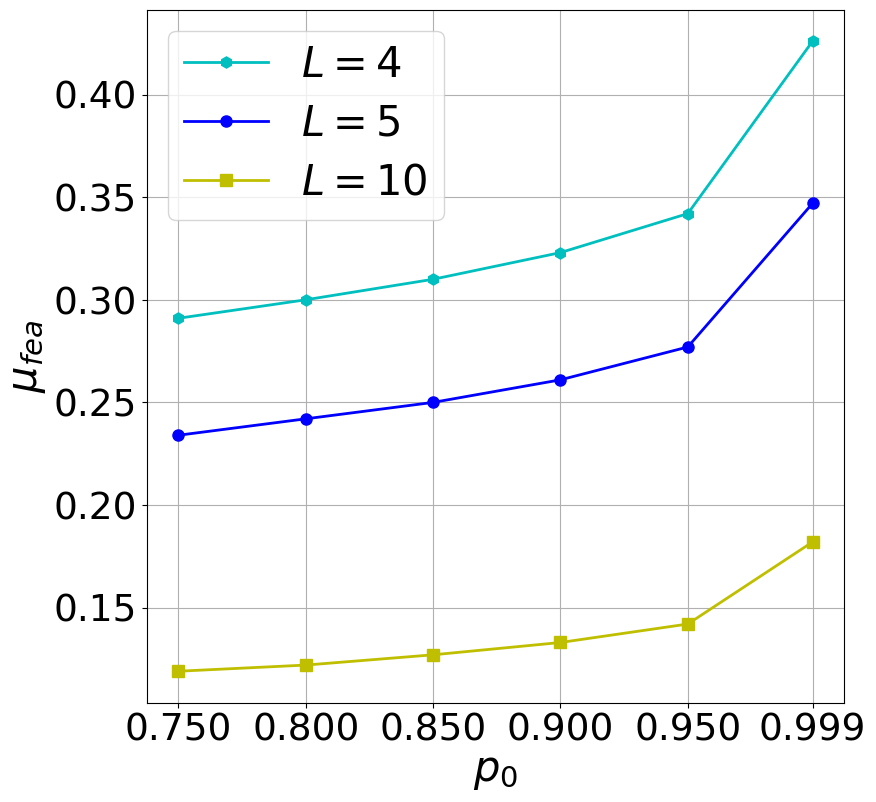}}\subfigure[Interval]{
\label{Fig.sub.3}
\includegraphics[height=2.5cm]{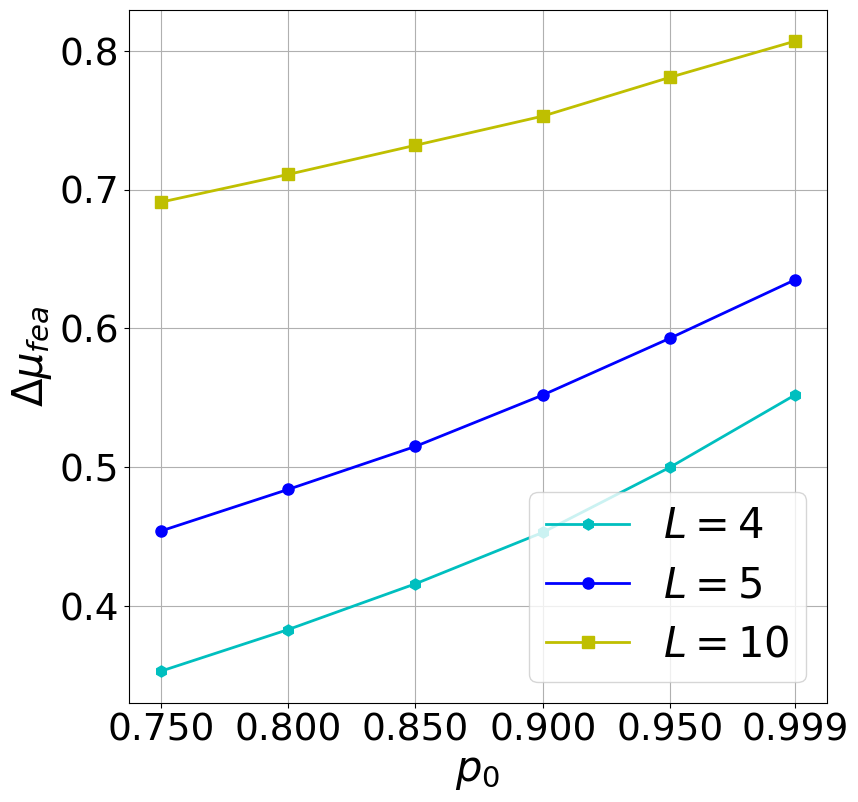}}
\caption{Comparisons of $\mu_{fea}$ with different $(L,p_0)$.}
\label{fig:9}
\end{figure}

% \begin{table}[htbp]
% \caption{Observability Probability Comparison of Different Methods}
% \begin{center}
% \begin{tabular}{c|c|c|c|c}
% \hline\hline
% \textbf{$(L,\beta)$}&\textbf{(5, 0.1)}&\textbf{(5, 10)}&\textbf{(10, 0.1)}&\textbf{(10, 10)} \\
% \hline
% SPM& 1  & 1 & 1 & 1   \\

% RSM& 0.0254 & 0.0729 & 0.2189 &0.0227   \\

% PSM&0.1152  & 0.2552 & 0.7303 &0.0227   \\

% OIDM& 0.1603 & 0.3541 & 1.1887 &0.0227   \\
% \hline\hline
% \end{tabular}
% \label{beta}
% \end{center}
% \end{table}

% p0与可行解mu_fea, L

% \begin{figure}[htbp]
% \centerline{\includegraphics[width=8.7cm]{pic4.png}}
% \caption{$\mu_{fea}-p_0$.}
% \label{fig:9}
% \end{figure}

% po对总、感知性能的影响
% Fig. 10, 

\section{Conclusion}

A stochastic sensor scheduling method guaranteeing probabilistic observability is proposed in this paper. Our OIDM enables a balance between sensing accuracy and power consumption with the collaboration of sensors and ECUs in the sensing process. Based on observability analysis and DDPG, OIDM provides new linear approximations of observability criteria, which guides the design of action space and improve learning efficiency. In future work, adding certain key deterministic sensor transmissions to stochastic scheduling can be considered to expand optimization space and guarantee observability strictly at the same time.

% we could further explore sensing, transmission and control co-design based on multiple ECUs' cooperation. 

\section*{Acknowledgment}

This work was supported in part by the NSF of China under Grants 61933009, 62025305, 92167205 and 61922058.

\addtolength{\textheight}{-12cm}   % This command serves to balance the column lengths
                                  % on the last page of the document manually. It shortens
                                  % the textheight of the last page by a suitable amount.
                                  % This command does not take effect until the next page
                                  % so it should come on the page before the last. Make
                                  % sure that you do not shorten the textheight too much.
\bibliographystyle{IEEEtran}
\bibliography{reference}
\end{document}